\documentclass{aa}
\usepackage[varg]{txfonts}
\bibpunct{(}{)}{;}{a}{}{,} 

\begin{document}
\title{A Steady-state Supersonic Downflow in the Transition Region above  a Sunspot Umbra}
\titlerunning{A Steady-state Supersonic Downflow in the TR above a Sunspot Umbra}
\author{Thomas Straus\inst{1} \and Bernhard Fleck\inst{2} \and Vincenzo Andretta\inst{1}}
\institute{INAF/Osservatorio Astronomico di Capodimonte, Via Moiariello 16, 80131 Napoli, Italy
\and ESA Science Operations Department, c/o NASA/GSFC, Greenbelt, MD 20771, USA}
\authorrunning{Straus et al.}
\date{Received 4 February 2015 / Accepted 15 July 2015}
\abstract{We investigate a small-scale ($\approx$ 1.5\,Mm along the slit), supersonic downflow of about 90\,km\,s$^{-1}$ in the transition region above the light-bridged sunspot umbra in AR\,11836. The observations were obtained with the Interface Region Spectrograph (IRIS) on 2013 September 2, from 16:40 to 17:59 UT. The downflow shows up as red-shifted ``satellite'' lines of the \ion{Si}{IV} and \ion{O}{IV} transition region lines and is remarkably steady over the observing period of nearly 80\,min. The downflow is not visible in the chromospheric lines, which only show an intensity enhancement at the location of the downflow. The density inferred from the line ratio of the red-shifted satellites of the \ion{O}{IV} lines ($N_\mathrm{e} = 10^{10.6\pm0.25}\,\mathrm{cm}^{-3}$) is only a factor 2 smaller than the one inferred from the main components ($N_\mathrm{e} = 10^{10.95\pm0.20}\,\mathrm{cm}^{-3}$). Consequently, this implies a substantial mass flux ($\approx 5 \times 10^{-7}\,\mathrm{g}\,\mathrm{cm}^{-2}\,\mathrm{s}^{-1}$), which would evacuate the overlying corona on time scales of the order of 10\,s. We interpret these findings as evidence of a stationary termination shock of a supersonic siphon flow  in a cool loop rooted in the central umbra of the spot.}
\maketitle

\section{Introduction}
Sunspots have been an area of intense research ever since Hale's 
\citeauthor{1908ApJ....28..315H}'s \citeyearpar{1908ApJ....28..315H} 
discovery of strong magnetic fields in these structures \citep[see reviews by, e.g.,][]{2011LRSP....8....4B,2011LRSP....8....3R}. Here we investigate a particular sunspot phenomenon: a small-scale ($\approx$\,1.5\,Mm along the slit), supersonic, relatively stable downflow of about 90\,km\,s$^{-1}$ in the transition region above a sunspot umbra, which was first noticed by \citet{2014ApJ...786..137T} in the same data set. 

A similar phenomenon has first been seen in data obtained during the 21 July 1975 flight of the NRL High Resolution Telescope and Spectrograph (HRTS) by \citet{1982SoPh...77...77D} and \citet{1982SoPh...81..253N}, and later in SOHO/SUMER observations by \citet{2001ApJ...552L..77B}  and \citet{2004ApJ...612.1193B}. Since HRTS provided only a snapshot, no inferences could be made about the temporal variability or stability of this feature. \citet{2001ApJ...552L..77B}, on the other hand, did have temporal information in the SUMER sit-and-stare time series they studied. In the 20-min time series they presented they found line profiles that were well represented by two Gaussian line components in the \ion{N}{V} 1242 {\AA} and \ion{O}{V} 629 {\AA} lines, with the main component showing prominent 3-min oscillations and the red-shifted component showing nearly the same high velocity ($\approx$\,90 km\,s$^{-1}$ in one data set) without oscillations in both emission lines. They introduced the term ``dual flow''. They did not discuss the origin of these dual flows in detail.  \citet{2004ApJ...612.1193B} extended their earlier work and 
found ``dual flows'' (i.e. line profiles in which two distinct velocities could be observed within the same resolution element) in 5 out of the 12 sunspots they studied. 

There have been numerous reports of supersonic downflows above sunspots following the initial HRTS publications by \citet{1982SoPh...77...77D} and \citet{1982SoPh...81..253N}. Most of them are based on observations with HRTS I and II and HRTS on Spacelab 2 \citep[e.g.][]{1990Ap&SS.170..135B,1991AdSpR..11..251B,1991ApJS...75.1337B,1988ApJ...334.1066K,1993SoPh..145..257K} and often showed multiple flow components within the $1\arcsec\times 1\arcsec$ resolution element. It is not clear, however, whether all these flows are akin to the "dual flows" of \citet{2001ApJ...552L..77B}.
\citet{1993ApJ...412..865G} on the other hand measured only weak, mostly subsonic downflows when analysing observations of eight sunspots with UVSP on the Solar Maximum Mission (SMM). He noted an interesting aspect in which the sunspots of his work differed from the ones observed by HRTS: the latter ones all had obvious light-bridges, while only one of the eight he observed with UVSP did. 
\citet{1988ApJ...334.1066K} did also speculate that the flows, which they found to have a typical size of 5",  may be associated with structures such as light bridges.

In addition to the two Brynildsen et al. papers cited above, there have been several other studies of transition region downflows above sunspots based on data from SUMER and CDS on SOHO \citep[e.g.][]{1998ApJ...502L..85B,2005ApJ...632.1196B}. These downflows, however, we believe are different from the phenomenon we are describing in this paper. They were not as strong, mostly subsonic or in any case below 35\,km\,s$^{-1}$, and more extended, often covering a good fraction of the penumbra. 

Supersonic downflows in the transition region above a sunspot have also been recently detected by \citet{2014ApJ...789L..42K} with the Interface Region Spectrograph \citep[IRIS;][]{2014SoPh..289.2733D}. These authors observed bursts of extremely broad line profiles in \ion{Mg}{II} h and k, \ion{C}{II} 1336\,\AA, \ion{Si}{IV} 1394\,\AA\ and 1403\,\AA, suggesting supersonic downflows of up to 200\,km\,s$^{-1}$ and weaker upflows that are associated with small-scale brightenings in sunspot umbrae and penumbrae. The broad line profiles do not have distinct velocity components as e.g. the multiple flows observed by \citet{1982SoPh...77...77D}, \citet{1982SoPh...81..253N}, \citet{1993SoPh..145..257K}, \citet{2001ApJ...552L..77B}, and \citet{2004ApJ...612.1193B} and they appear rather intermittently in short bursts lasting about 20\,s. These characteristics are very different from the rather steady, ``mono-chromatic'' downflow discussed in this paper,  which reveals itself as Gaussian ``satellite'' lines in the spectra of \ion{Si}{IV} and \ion{O}{IV}, well separated from the main components. Interestingly, such a well-separated component is also visible in the $\lambda$--t--plots displayed in Figure 3 of \citet{2014ApJ...789L..42K} at about 100 km\,s$^{-1}$ red-shift, but not discussed in the paper.

In Section 2 we will describe the observations. The analysis and results are presented in Section 3 and discussed in Section 4, before concluding in Section 5.

\section{Observations}

\begin{figure*}
\centering
   \includegraphics[width=17cm]{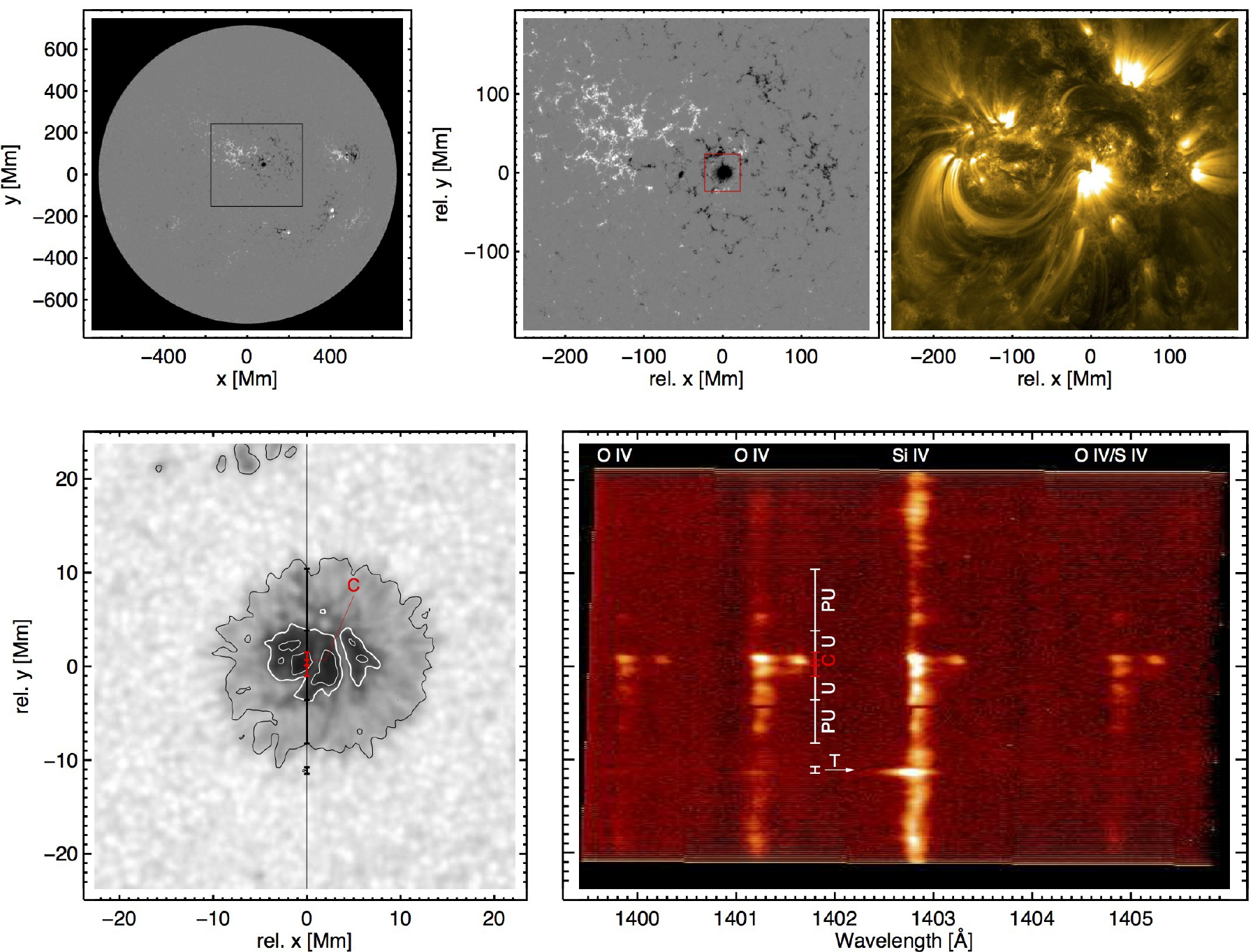}
     \caption{Upper panels (context images of AR 11836): full disk (left) and cut-out (middle) HMI magnetogram and corresponding AIA 171\,\AA\ image (right). Lower panels: zoom of HMI continuum image on the area marked by the red box in the upper panel, with the IRIS slit marked at x=0 (left), and the corresponding spectra in the \ion{Si}{IV} 1403\,\AA\ window averaged over 1\,minute (right panel). All data were taken on 2013 Sep 2, 17:14 UT. Regions where the slit crosses the penumbra (PU) and umbra (U) are marked. The central part of the umbra where the \ion{Si}{IV} and \ion{O}{IV} lines show satellites is marked in red. In the northern part the satellites are strongest, the center of this region is labelled ``C'' and used as the reference point of the satellite feature in the following. The prominent upflow labeled "T" is due to a short-lived transient.}
     \label{spot}
\end{figure*}

For this study we used the same data set as was used by \citet{2014ApJ...786..137T} for their study of the propagation behaviour of shock waves in a sunspot. The data were obtained with the IRIS in sit-and-stare mode on 2013 September 2 from 16:39:35 to 17:58:45 UT. The target was the sunspot in AR\,11836 near disk center at x=106\arcsec, y=58\arcsec (coordinates of the center of the slit at the beginning of the observations). Slit length and width were 68\arcsec and 0\farcs 166, respectively. The pixel size along the slit is 0\farcs 166. The data were acquired by binning the FUV data in the wavelength direction by a factor of 2, and thus have a dispersion of $\approx$ 25\,m\AA\ per pixel (about 5\,km\,s$^{-1}$ at 1400\,\AA) in both the NUV and FUV channels. The cadence of the spectral data was 3\,s, with exposure times of 2\,s. The observations comprise nine spectral windows: \ion{C}{II} 1336\,\AA,  \ion{Fe}{XII} 1349\,\AA,  \ion{Cl}{I} 1352\,\AA,  \ion{O}{I} 1356\,\AA,  \ion{Si}{IV} 1394\,\AA,  \ion{Si}{IV} 1403\,\AA, NUV at 2786\,\AA\ and 2831\,\AA, and  \ion{Mg}{II} h and k 2796\,\AA. 
The spectral window containing the \ion{Si}{IV} 1403\,\AA\ line also includes the \ion{O}{IV} 1400\,\AA, 1401\,\AA, and 1405\,\AA\ lines, the last one blended with a \ion{S}{IV} line. Parallel to the spectral data slit-jaw images were taken in three filters (2796\,\AA, 1400\,\AA\ and 1330\,\AA) at a 12\,s cadence. The pixel size of the slit-jaw images is 0\farcs 166$\times$0\farcs 166. We used level-2 data from the IRIS archive at Lockheed-Martin Solar and Astrophysics Laboratory (LMSAL), i.e. data which have already been corrected for dark current, flat field and geometrical distortions. Below we also discuss the line ratios of the \ion{Si}{IV} doublet and the \ion{O}{IV} density sensitive lines. For that purpose we applied the current estimate of the wavelength-dependent effective area as provided by the Solarsoft routine \texttt{iris\_get\_response.pro}, using version 3 of the calibration, released on 2015 April 1.

\begin{figure*}
\centering
   \includegraphics[width=17cm]{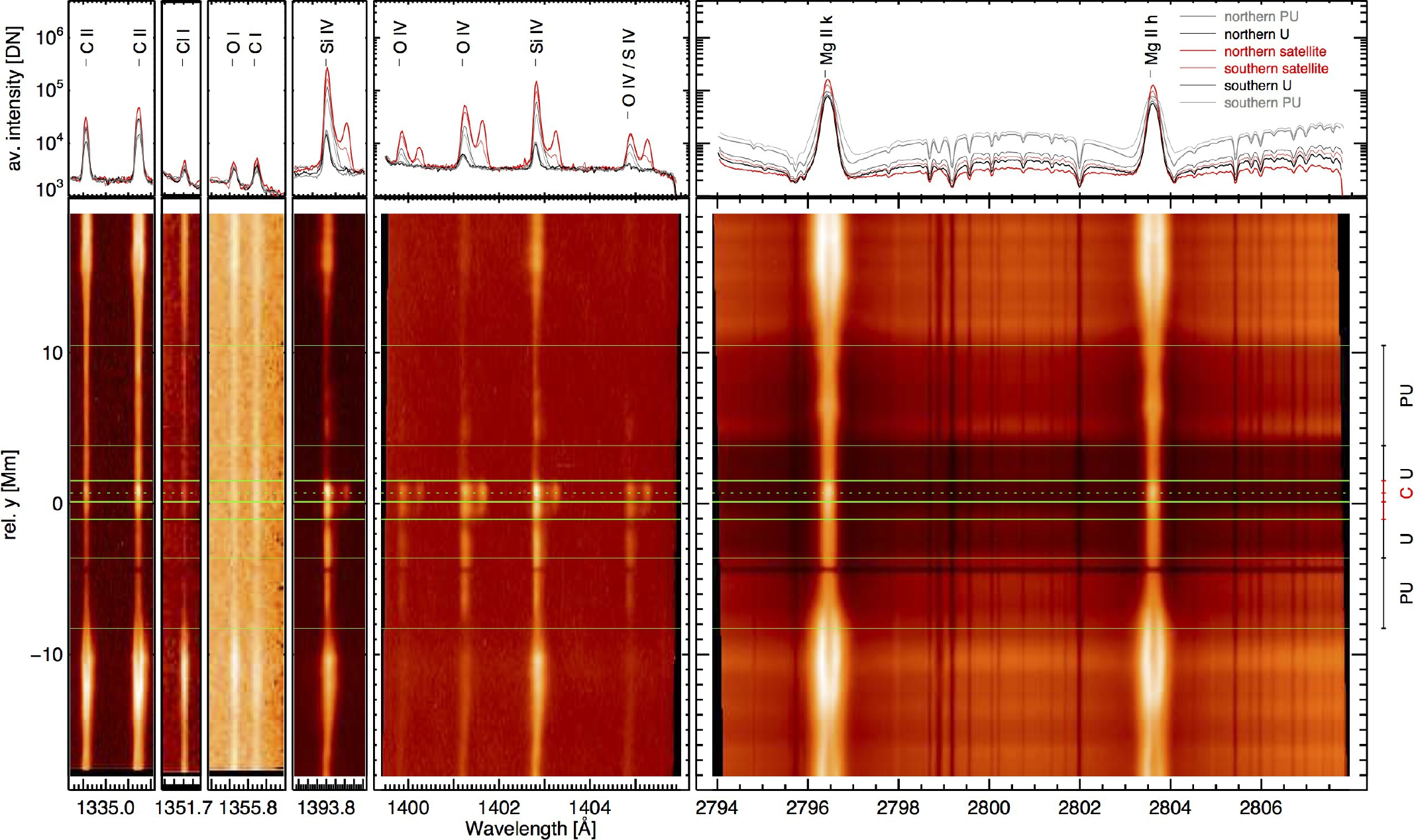}
     \caption{Spectra averaged over the whole time series as function of wavelength and position along the slit (lower panels, logarithmic color scale) and further averaged over the different regions (upper panels). The marks of different regions in the lower row correspond to those given in Fig.~\ref{spot}.}
     \label{avspectra}
\end{figure*}

For context and to gain some information about the higher temperature plasma and the magnetic field structure we complemented the IRIS observations with data from the Atmospheric Imaging Assembly \citep[AIA;][]{2012SoPh..275...17L} and the Helioseismic and Magnetic Imager \citep[HMI;][]{2012SoPh..275..207S} on the Solar Dynamics Observatory \citep[SDO;][]{2012SoPh..275....3P}. To allow a comparison with the IRIS observations, the AIA and HMI data were remapped to the IRIS data, both spatially and temporally, using metadata in the FITS headers and the map routines of SolarSoft. No additional roll angle correction has been applied. A secondary step in SDO-IRIS cross-alignment has been achieved by comparing the HMI white-light images and the near continuum data from the 2831\,\AA\ spectral window of IRIS (wavelength index 15). The different IRIS spectral windows have been cross-aligned using the fiducial lines. No further inter-instrument alignment between HMI and the different AIA channels has been applied. In a final step, temporal jitter and drift of IRIS along the slit has been removed from the spectral data by cross-correlation techniques applied to the {\bf near} continuum intensity in the 2831\,\AA\ spectral window. The totally accrued drift during the observations was less than 3.5 IRIS pixels or, equivalently, one HMI pixel (0\farcs 5). Pointing excursions of IRIS perpendicular to the slit cannot be corrected for. Altogether, we believe that the cross-alignment between the IRIS and SDO data is within one HMI pixel. 

\section{Observational results}\label{sec:observations}

We start with a summary of the characteristics of the observed supersonic downflow. The various findings are discussed in more detail below. Note that throughout this paper we adhere to the convention that positive velocities represent blueshifts. The downflow
\begin{itemize}
\item shows up as red-shifted ``satellite'' lines of the \ion{Si}{IV} and \ion{O}{IV} transition region lines, completely separated from the main components by about 90\,km\,s$^{-1}$ (Fig.~\ref{spot} -- \ref{flows}), 
\item is very localised and small-scale ($\approx$ 1.5\,Mm along the slit; Fig.~\ref{spot}),
\item occurs in the center of a sunspot umbra, which has one prominent and two smaller light-bridges (Fig.~\ref{spot}),
\item is a very steady-state flow both in line intensity and velocity (Fig.~\ref{flows}), lasting at least as long as the observations, with surprisingly little variation for the transition region, which usually is highly dynamic (cf. Kleint et al., 2014),
\item does not participate in the 3-min shock wave dynamics, which dominate the main components (Fig.~\ref{flows}), and shows only very weak, anti-correlated intensity signatures (Fig.~\ref{shocks}),
\item is comparatively more pronounced (at about 60\% $\pm$ 20\% of the main component) in the  \ion{O}{IV} lines than in the \ion{Si}{IV} doublet ($\sim$ 20\% $\pm$ 10\% of the main component),
\item has a density of about 4$\times 10^{10}$\,cm$^{-3}$, with little temporal and spatial variations, and accordingly a mass flux of $\approx \mathbf{5} \times 10^{-7}$\,g cm$^{-2}$\,s$^{-1}$,
\item has a density about a factor $\mathbf{\approx 2}$ lower than the one inferred from the main component (Fig.~\ref{density}),
\item is visible only in the \ion{Si}{IV} and \ion{O}{IV} transition region lines, with no velocity signature in the chromospheric lines \ion{Mg}{II} h and k or \ion{C}{II} (Fig.~\ref{avspectra}),
\item is co-spatial with an increased intensity in those same chromospheric lines, suggesting related enhanced heating  (Fig.~\ref{avspectra}),
\item is not characterised by distinctly peculiar line profiles; in particular the profiles of the \ion{Si}{IV} and \ion{O}{IV} satellite lines are nearly Gaussian with widths which are about the same as that of the main components, although the width of the \ion{Si}{IV} satellite lines tends to be larger than the main component, by about 40\% with large uncertainties (Fig.~\ref{linewidth}).

\end{itemize}

\subsection{Context of observation}\label{sec:context}

Figure~\ref{spot} provides an overview of the region studied in this paper. The upper left shows a HMI full disk magnetogram, and the upper right a cut-out HMI magnetogram and AIA 171\,\AA\  intensity image of AR 11836. The lower left shows an even more zoomed-in HMI intensity image of the central sunspot of AR 11836. The position of the IRIS slit is indicated (x = 0). The lower right shows a 1-min average of the spectral window containing the \ion{Si}{IV} 1403\,\AA\ line. The red marks indicate the region (about 2.8~Mm wide) in the umbra where the \ion{Si}{IV}  and \ion{O}{IV} lines show satellites. In the northern part the satellites are  strongest; the size of this northern part is about 1.5\,Mm. The center of this region is labelled ``C'' and used as the reference point of the satellite feature in the following. 

The same data set has been studied and described by \citet{2014ApJ...786..137T} and \citet{2015ApJ...798..136Y}, who investigated the strong 3-min umbral oscillations found in this spot. It is interesting to note that point ``C'' seems to coincide both with the apparent source of the umbral running waves studied by \citet{2015ApJ...798..136Y} and with the location of the most intense magnetic field in the sunspot.

As can be seen in the HMI images, the spot of AR 11836 is an isolated, small, round spot of leading polarity. There is no following spot, only active region plage. The spot is surrounded by a large area with network field of the same polarity. Opposite polarity can be found in onl    y a few locations at the south and north-east of the spot. The loops visible in AIA 171\,\AA\  are inclined and connect the spot to plage of opposite polarity following the spot. The loops seem to be anchored in the dark north-west quadrant, whereas the rest of the sunspot displays a bright plume in the AIA 171\,\AA\  images. 

The sunspot shows an irregular umbra, split up in two pieces by a prominent light-bridge in the western part. The bigger, eastern part is further split into three pieces by two smaller, less prominent light-bridges. According to \citet{1993ApJ...412..865G} it might be an important aspect for the presence of a supersonic downflow that this sunspot had a light-bridge. The slit of the IRIS spectrograph crossed the eastern part centrally in the north-south direction.

\subsection{Average properties}\label{sec:averages}

The \ion{Si}{IV} and  \ion{O}{IV} lines form at approximately 70\,000\,K and 150\,000\,K, respectively. 
In both lines, the supersonic downflow is stable enough to be prominent  in the spectra averaged over the whole 80\,minutes of the observations  (see Fig.~\ref{avspectra}). The strong northern satellite region is the darkest  part in the ``pseudo-continuum'' between the \ion{Mg}{II} h and k lines  and shows at the same time the brightest intensity in all emission  lines. There is no hint of any red-shifted components in the  chromospheric spectral bands, in particular also not in the strong  \ion{Mg}{II} h and k and \ion{C}{II} 1336\,\AA\ lines. 

In order to proceed with a quantitative analysis of the line profiles, we determined the main line parameters, namely the peak value, $I_\mathrm{max}$, its position, $\lambda_\mathrm{max}$, and the full width at half maximum, FWHM, by means of a parabolic fit to the central three pixels of each profile. Since the core of the \ion{Si}{IV} and \ion{O}{IV} lines is only 3 or 4 pixels wide in this data set, a fit to a larger number of pixels is normally not required. At the extrema of shock maneuvres, double-peaked profiles may appear: the fit procedure then identifies the strongest peak. If needed, the total line intensity, $I_\mathrm{tot}$, is computed like in the case of Gaussian profiles as the peak value times the line width: $I_\mathrm{tot}=0.5\:\sqrt(\pi/\log 2)\times I_\mathrm{max}\times\mathrm{FWHM}=1.064467\times I_\mathrm{max}\times\mathrm{FWHM}$.  We have compared this procedure with a more standard Gaussian fit, verifying that the agreement is excellent, especially for the more prominent lines. The Gaussian fit procedure, however, tends to fail more often in the case of weak profiles.  The double peaks at the extrema of shock maneuvres are also less clearly identified.  Finally, Gaussian fits assume symmetric profiles, while parabolic fits do not make this assumption.

\begin{figure*}
\sidecaption
   \includegraphics[width=12cm]{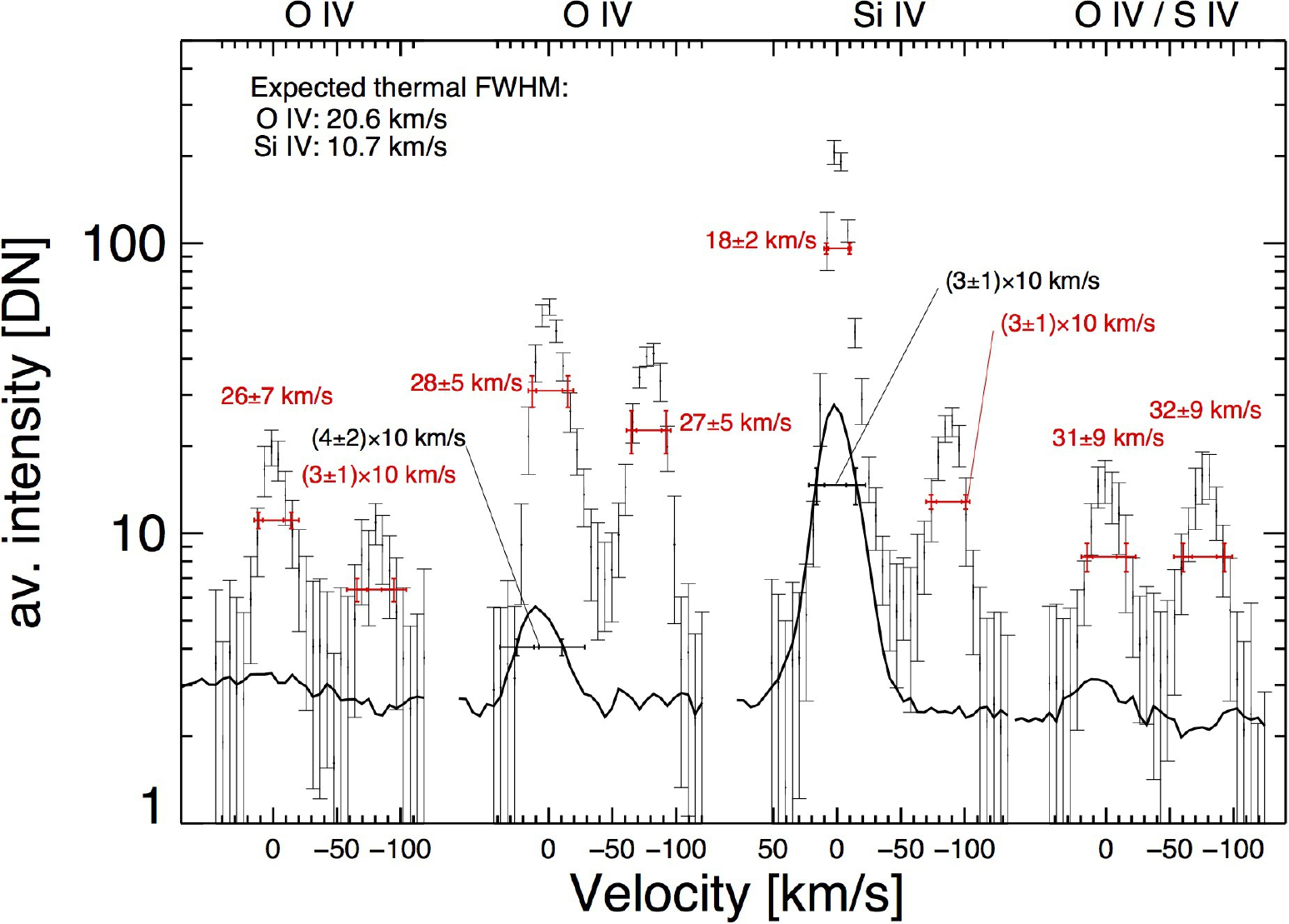}
     \caption{Symbols: 1-min average line profiles at point ``C'' with error bars (standard deviation of the twenty individual measurements), taken 5 minutes before the reference spectrum in Fig. \ref{spot} (17:09 UT). At this time, the satellite lines in \ion{O}{IV} are particularly strong. Solid line: average spectrum outside the spot. The main and satellite lines have all similar line widths (FWHM$\approx 30\,$km\,s$^{-1}$), with the exception of \ion{Si}{IV}\,1403\,\AA\ which has a significantly narrower main component. The satellites are less pronounced in \ion{Si}{IV} than in \ion{O}{IV}.} 
     \label{linewidth}
\end{figure*}

\begin{figure*}
\centering
   \includegraphics[width=17cm]{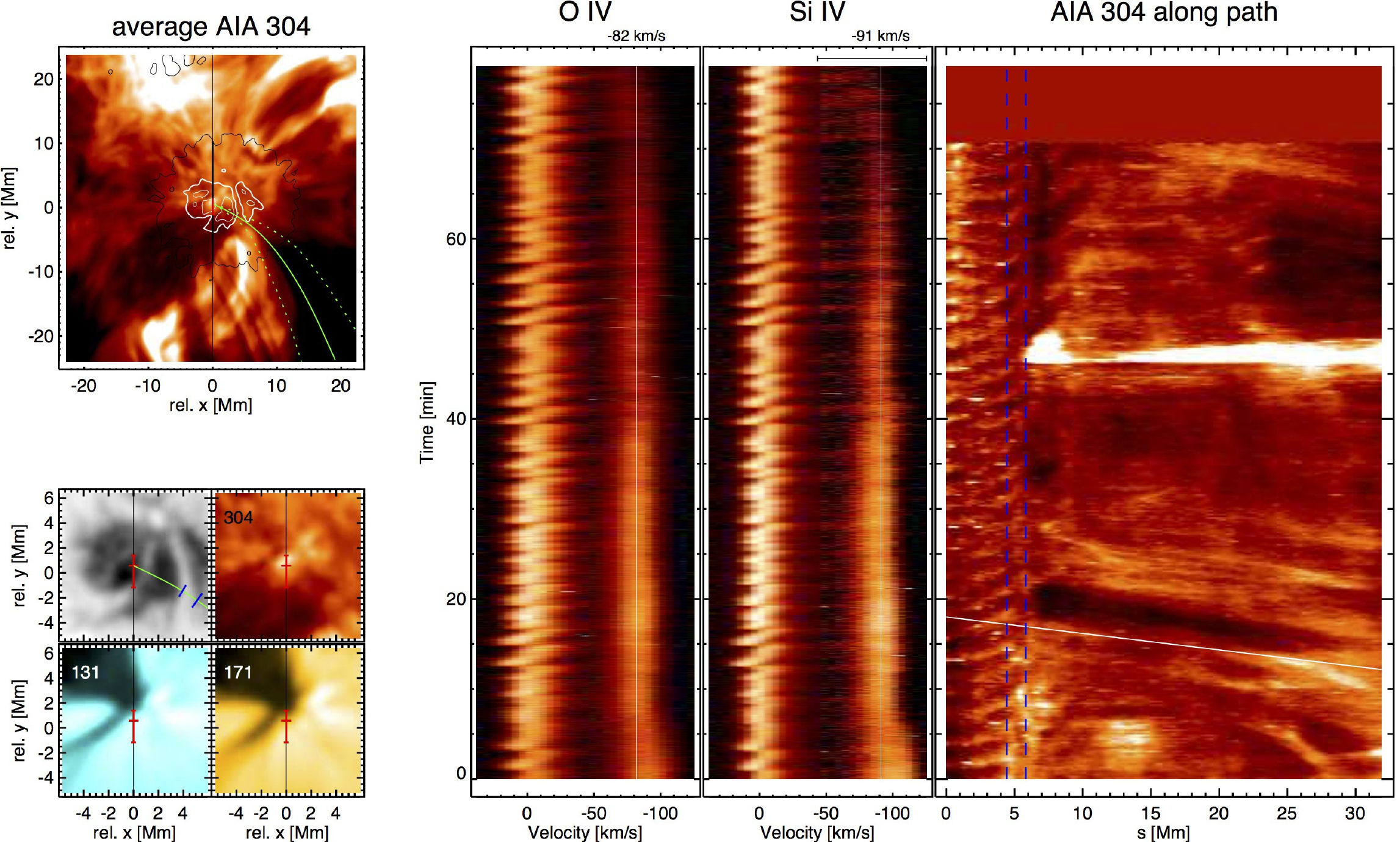}
     \caption{Upper left panel: temporal average of AIA 304\,\AA\ intensity, with HMI continuum intensity contours from Fig.~\ref{spot} overlayed. The green solid and dashed lines indicate a path of inflows and the area of integration used for the time-distance plot in the right panel. Lower left panels: temporal average images of HMI continuum intensity and three AIA channels. The red marks indicate the satellite region and point ``C'' as in Fig.~\ref{spot}. Central panels: temporal evolution of the \ion{O}{IV} and \ion{Si}{IV} lines averaged over 13 pixels in the northern satellite region centered on point ``C''. The region of the satellite in \ion{Si}{IV}, marked by the bar on top, has been contrast-enhanced to better show the weak connection to the main component's shock maneuvres. Right panel: relative AIA 304\,\AA\ intensity fluctuations integrated across the path marked by the green lines in the context image in the upper left. The colortable maps variations from -30\% to +30\%. Coronal rain is visible with plane-of-sky velocities of the order of the downflow velocity of the satellite (90 km\,s$^{-1}$, marked by the white line). However, these inflows do not seem to extend to the position of the satellite feature in the umbra, but appear to end at the light-bridge as marked by the vertical, blue lines in the time-distance plot and the corresponding marks on the path in the HMI continuum context image.}
     \label{flows}
\end{figure*}

The  average properties of the line parameters thus estimated were then determined from the histograms  computed over 13  pixels centered on point ``C'' along the slit, i.e. the region within the two marks nearest to point ``C'' in Fig.~\ref{spot} (the northern satellite region). A Gaussian fit to the peak of the histogram yields both the average value and the standard deviation. We estimated the averages both over the full time span and for the first 30  minutes of the observing run only. We  computed the average over  the first 30 minutes because after that time the red-shifted components become fainter. Moreover, after about 30 minutes, the mean position  along the slit of the downdraft region starts drifting south with respect to point ``C'' by several pixels. The average values do not change much in either case: for  the \ion{Si}{IV} line the mean shift between the two components is 91 $\pm$ 8  km s$^{-1}$, while the average value for the first 30 minutes is 94 $\pm$ 6  km s$^{-1}$; the corresponding values for the \ion{O}{IV} lines are 82 $\pm$ 8 km s$^{-1}$, while  the average for the first 30 minutes is 84 $\pm$ 6 km s$^{-1}$. 
In the following we will always refer to averages computed for the first 30 minutes, unless explicitly stated otherwise.

The satellite lines are most pronounced in \ion{O}{IV}, with the total intensity being {\bf on} average approximately 60$\pm$15\% of the main component. The average \ion{Si}{IV} satellite line is about  20$\pm$10\% of the total intensity of the main component. Variations in the relative intensity of the satellite to the main component in different \ion{O}{IV} lines can be interpreted as due to density   variations (see discussion of Sec.~\ref{sec:plasma:density}).  Since we do not have estimates of plasma density in the region where the \ion{Si}{IV} lines form, we are unable to offer a simple explanation of the different behaviour of this ratio between satellite and main component to the \ion{O}{IV} lines. This piece of information may be useful as constraints to more detailed models of this phenomenon. However, we note that already \citet{1982SoPh...81..253N}, using a larger set of lines, estimated that the shape of the differential emission measure of the supersonic downflows is drastically different from the curve corresponding to the surrounding umbra. In particular, the differential emission measure they found in the downflows (their Fig. 8) is similar to the one they find in the sunspot umbra for $\log T > 5$, while it continues to drop towards lower temperatures, as opposed to the increase after  that turn-around point seen in the umbra. This implies that the intensity of the \ion{O}{IV} in downflows should be comparable to the one observed in the surrounding umbra, while the \ion{Si}{IV} lines intensities should be strongly suppressed, as observed.

We performed a similar analysis on the southern part of the satellite region.  There the satellite lines are weaker compared to the main component: the \ion{O}{IV} satellite lines are {\bf on} average only about 10\% of the main component, while the \ion{Si}{IV} satellites are 5\% or less.  However, the velocity shifts remain as stable and comparable in magnitude to the values measured around point ``C'', although slightly smaller {\bf on} average: $\approx70$ km s$^{-1}$ for the \ion{O}{IV} lines, and $\approx80$ km s$^{-1}$ for the \ion{Si}{IV} lines.

\subsection{Plasma diagnostics}\label{sec:plasma}

For the remainder of this section, except when noted otherwise, we analysed line profiles calibrated according to the procedure described in IRIS Technical Notes \#1 and \#24\footnote{\texttt{http://iris.lmsal.com/documents.html}}, and adopted the most recent effective areas for the IRIS spectrograph as provided by the SolarSoft routine \texttt{iris\_get\_response.pro} (version 3, 2015 April 1). 

\subsubsection{Line optical thickness}

  In the plage regions outside the sunspot, the profiles of the \ion{C}{ii} lines at 1334.5\,\AA\  and 1335.7\,\AA\ display an evident self-reversal, a clear signature of a significant optical thickness.  Even in the penumbra the ratio of the total intensities of the two components, in contrast with the optical thin limit of 1:2, is around 1:1.4, a value close to the observed value in quiescent regions \citep{2003ApJ...597.1158J} and to the computed values in the reference quiescent chromosphere by \cite{2008ApJS..175..229A}. 

On the other hand, the profiles in the region around point ``C'' are nearly Gaussian. Nevertheless, the line widths of the \ion{C}{ii} 1334.5\,\AA\ are {\bf on} average larger than those of the \ion{C}{ii} 1335.7\,\AA\ line in the northern satellite region ($1.3\pm0.2$) while at the same time the ratio of the total line intensities is consistent with the optical thin value ($2.0\pm0.3$)\footnote{The 1335.7~\AA\ line is in reality composed of two lines, at 1335.71~\AA\ and 1335.66~\AA; this latter line, however, is much weaker and does not therefore affect significantly the width of the observed profile.}. These two measurements together suggest that the \ion{C}{ii} doublet in that region is affected by a modest optical thickness which is possibly smaller than in the quiet Sun, and that it is most likely being formed in the effectively optically thin regime: the line profiles become broader before the total line intensity is reduced by the destruction of photons due to collisions in the multiple scattering process.These measurements are consistent with earlier estimates \citep[e.g.][]{1978ApJ...226..687J,1979MNRAS.187..463B,2009A&A...505..307T} suggesting a lower opacity of the chromosphere above  sunspot umbrae.

The average line profiles of the \ion{Si}{IV} 1393.8\,\AA\  and 1402.8\,\AA\ lines do not show any sign of self-reversal or of significant departures from Gaussian in the line core in any region crossed by the slit (Fig.~\ref{avspectra}); we could not find any clearly non-Gaussian profiles that could be attributed to optical thickness effects in the data we browsed, although we could not check all the individual profiles in the data set.  

A first analysis of the intensity ratio in the northern satellite region around point ``C'' of the main components of the two \ion{Si}{IV} lines showed nevertheless the mean ratio of the 1394~\AA\ to the 1403~\AA\ to be around 1.4, a value which is significantly lower than the optical thin line ratio of 2.  During the course of the revision of this paper, however, an updated, post-launch radiometric calibration of IRIS was released.  With this latest calibration, the ratio of the main components of the \ion{Si}{IV} 1393.8 and 1402.8~\AA\ lines became $1.9\pm0.1$, much closer to the optically thin ratio. The ratio of the widths of the main component of the \ion{Si}{IV} 1402.8\,\AA\  line to the 1393.8\,\AA\ line is on average $1.04\pm0.08$ in the same region, a value close to the value of 1.07 given by \cite{2004ApJ...600.1061D} for the quiet Sun.

If we regard the results of the analysis of the \ion{C}{ii} lines above as relevant also for the main \ion{Si}{IV} lines, i.e.\ that the optical thickness in point ``C'' is even smaller than in the average quiet Sun, we can also cite previous studies carried out in quiescent regions.  In particular, the analysis of \cite{1979MNRAS.187....9R} on the limb brightening of the line intensity of \ion{Si}{IV} 1393.8\,\AA\ in OSO-8 data indicates an optical depth of $0.22\pm0.16$ in the quiet Sun, in contrast with the much larger optical thickness of the \ion{C}{ii} 1334.5\,\AA\  and 1335.7\,\AA\ doublet estimated by \cite{2008ApJS..175..229A}. Moreover, the \ion{Si}{IV} lines 1393.8\,\AA\  and 1402.8\,\AA\ lines are not listed in Table~3 of \cite{2004ApJ...600.1061D} among the lines with measurable optical thickness in the FUV spectral range observed by SOHO/SUMER.

On the other hand, 
the ratio of intensities of the satellite lines with the same post-launch  calibration has the lower value of $1.6\pm0.3$. The ratio of  widths is $1.1\pm0.3$. Moreover, 
the \ion{Si}{IV} satellite lines are clearly  broader than their main components: the ratio of the FWHM of the satellite lines to the corresponding main component is $1.35\pm0.4$ (see also Fig.~\ref{linewidth}). This is in contrast with  the observation that the widths of the \ion{O}{IV} satellite lines are approximately equal to those of their main components (ratio: $0.95\pm0.15$).  Since the peak intensities of the \ion{Si}{IV} and \ion{O}{IV} satellite lines are comparable (see again Fig.~\ref{linewidth}), it is unlikely that a bias due to a possible instrumental background or solar continuum could be invoked to explain these measurements.

It would then be  tempting to interpret the larger width of the \ion{Si}{IV} satellite lines as  an optically thickness effect.  Following the approach of \citet{2004ApJ...600.1061D},  this increased width would imply an optical thickness of the order of  unity or larger (see for instance their Fig.~3). An optical thickness of the order of unity in the \ion{Si}{IV} 1394~\AA\ line, with the  densities of the order of $10^{10}$--$10^{11}$ cm$^{-3}$ as inferred in the following Sec.~\ref{sec:plasma:density} implies a geometrical thickness of the emitting  plasma of at least a few hundreds of kilometers. Whether optical thickness effects alone can explain the broader \ion{Si}{IV} satellite lines (as well as their ratio) will need to be verified by more detailed models.

\begin{figure*}
\centering
   \includegraphics[width=17cm]{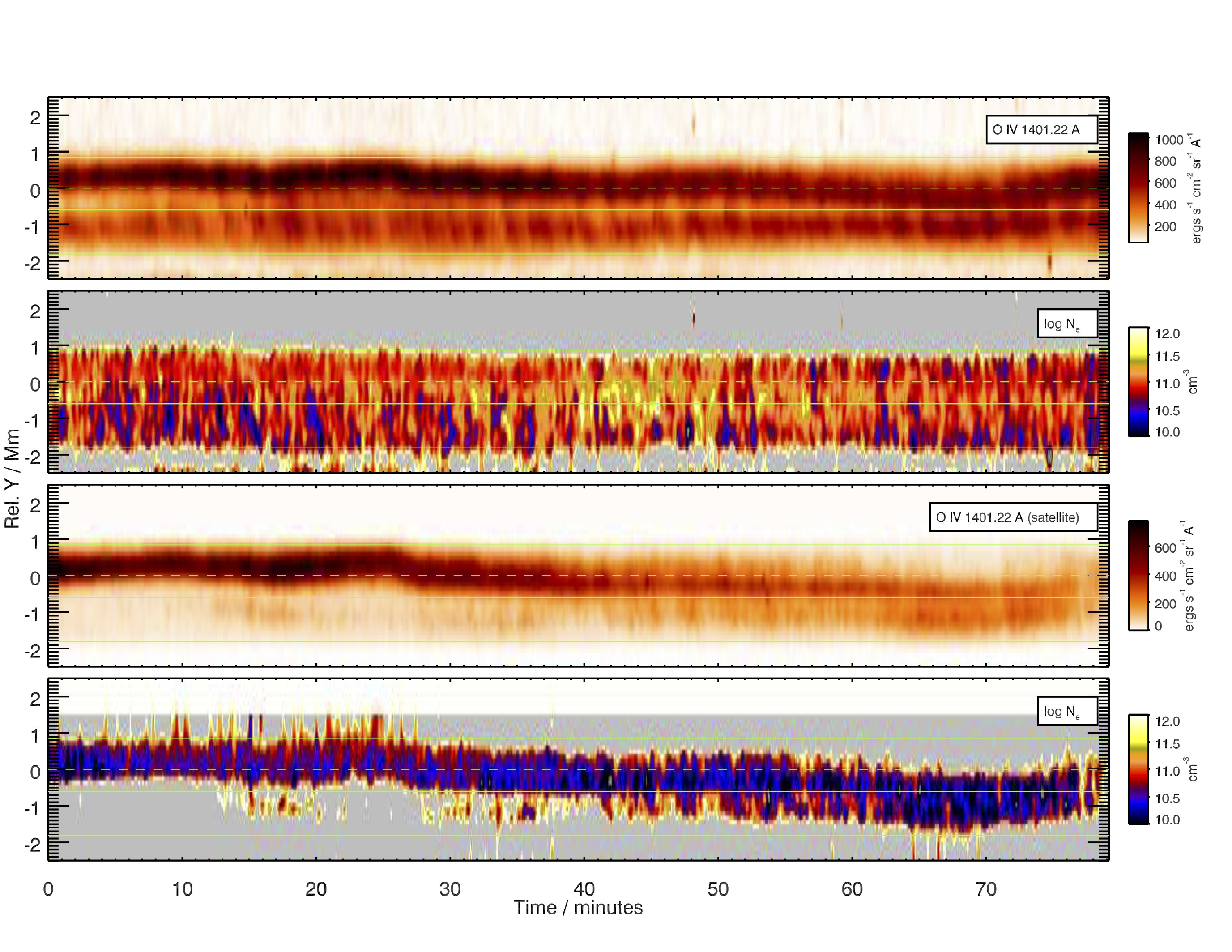}
   \caption{Variation of the intensity of the main component of the \ion{O}{IV} 1401\,\AA\  line in the region that showed a supersonic downflow (first panel from the top), together with density inferred from the ratio of the \ion{O}{IV} 1400\,\AA\  and 1401\,\AA\ lines (second panel). The variation of the intensity of the satellite line at 1401\AA\ (third panel) and the density from the ratio of the satellite lines at 1400\,\AA\  and 1401\,\AA\ are shown in the third and fourth panel, respectively. The horizontal lines mark the boundaries of the regions given in Fig.~\ref{spot}; the dashed line marks point ``C''.}
     \label{density}
\end{figure*}

\subsubsection{Electron density}\label{sec:plasma:density}

  The intercombination multiplet of \ion{O}{IV} lines at 1397.20\,\AA, 1399.77\,\AA, 1401.16\,\AA, 1404.81\,\AA, and 1407.39\,\AA\ provides a well known set of density-sensitive pairs \citep[e.g.:][]{1992SoPh..138..283D}; the lines at 1397.20\,\AA\ and 1407.39\,\AA\ are however outside the spectral window of this data set, while the line at 1404.81\,\AA\ is blended with a \ion{S}{iv} line. We thus considered the ratio between the \ion{O}{IV} 1399.77\,\AA\ and 1401.16\,\AA\ lines as density diagnostics.  We adopted the CHIANTI database of atomic data version 7.1.4 \citep{1997A&AS..125..149D,2013ApJ...763...86L}, assuming a Maxwellian distribution.  It should be noted that \cite{2014ApJ...780L..12D} recently studied the effects of non-Maxwellian distributions of electron energies on both the \ion{Si}{IV} and \ion{O}{IV} multiplets. We feel however that it would be unlikely that the stationary flows we are examining could sustain substantial departures from equilibrium energy distributions.

\begin{figure*}
\centering
   \includegraphics[width=17cm]{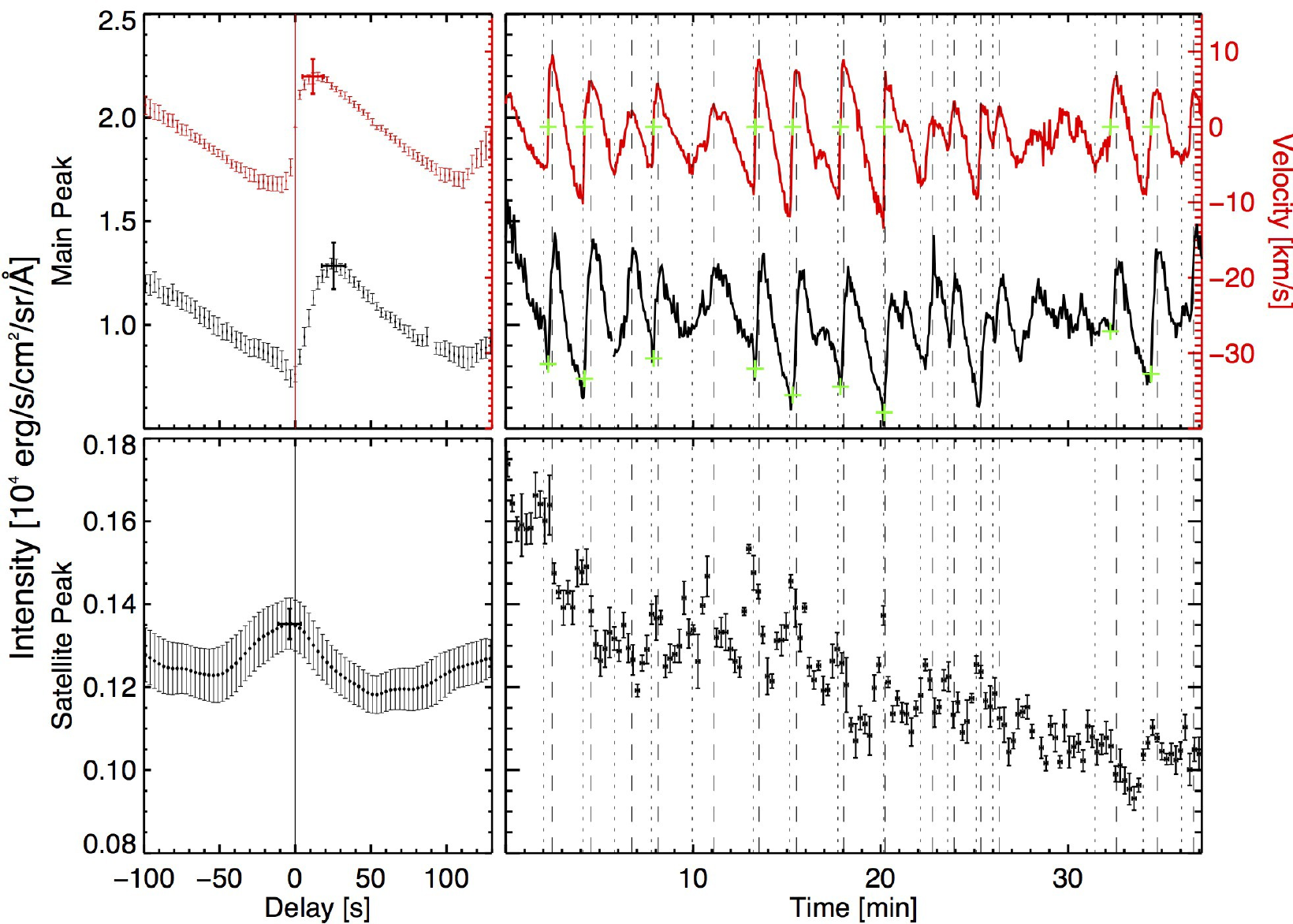}
     \caption{Right panels: temporal evolution of \ion{Si}{IV} line parameters averaged over the region where the line shows a satellite (black: intensity; red: velocity). Dotted and dashed vertical lines mark the beginning and end of shock front passages, respectively. Positive velocities correspond to blue shifts. The times of zero velocity in the shock fronts are marked by green crosses. The main component's peak intensity shows a minimum at these times. The lower right panel shows the temporal evolution of the satellite's peak intensity. Left panels: temporal evolution of the average shock front of the 9 shocks marked by crosses in the upper right panel; time is given relative to the zero-crossing times of the main component's Doppler-shift. The intensities of the \ion{Si}{IV} satellite and main component are anti-correlated.}
     \label{shocks}
\end{figure*}

The time variation of densities inferred from this ratio is shown in Fig.~\ref{density}, alongside the variation of the line intensities of both the main and the satellite components. To improve the signal-to-noise ratio, the line intensities were smoothed using a bi-Gaussian kernel with FWHM (full width at half maximum)  of  3 pixels along the slit and of 5 spectra in the time domain, corresponding to $0.35$\,Mm and $15$\,s, respectively.
Note that smoothing data does not resolve the issue of the presence of possible errors in determining the background continuum which is to be subtracted from the line intensities before computing the line ratio. Since the \ion{O}{IV} 1400\,\AA\ line intensity is {\bf on} average about 1/4 of the \ion{O}{IV} 1401\,\AA\ line in the region showing the supersonic downflow, a systematic error in the background correction will affect the former line more than the latter. The result would be a systematic bias in the determination of the density.  For instance, a residual background intensity in the \ion{O}{IV} lines would explain why the line ratio and therefore the density appears to increase in those regions where the lines are weaker (e.g. at the edge of the region where the satellite lines are visible).

We examined the statistical distribution of densities derived from the ratio of the main components of the two \ion{O}{IV} lines for the full spatial and temporal range over which the downflow is observable in this data set, i.e. for the full 80 minutes of the observing run and within 2.5~Mm from point ``C'' (Fig.~\ref{density}). 
We obtained a mean value of $\log N_\mathrm{e} = 10.95\pm0.20$ (1\,$\sigma$ uncertainty, $N_\mathrm{e}$ in cm$^{-3}$), while the densities from the satellite lines are $\log N_\mathrm{e} = 10.6\pm0.25$.

\subsection{Signatures of shock dynamics}\label{sec:shocks}

The temporal fluctuations of the \ion{Si}{IV} and  \ion{O}{IV} lines show  their main components with prominent shock wave passing maneuvres and  rather stationary supersonic downflow satellite lines. The satellite  lines are {\it not} participating in the shock wave maneuvres, but in  both lines the satellite intensity is slightly linked to the Doppler  shift of the main component, showing a slightly brighter satellite  feature when the main component passes through zero velocity in the steep rise of the shock front from maximum red-shift to blue-shift (Fig.~\ref{flows} and  \ref{shocks}). This suggests that the plasma of the downflow is to a first approximation dynamically decoupled from the plasma that forms the main components.  One could speculate about a plume of transition region temperature  plasma above the actual transition region. This would be a different  scenario from the cartoon presented in Fig.~14 of  \citet{1982SoPh...81..253N}. However, we detect subtle signatures of  a connection between the main and red-shifted component, as we discuss below.

In Fig.~\ref{shocks} the main line parameters determined as described in Sec.~\ref{sec:averages} are shown after averaging over all 13 pixels in the northern part of the downflow region. The shock-like temporal fluctuations are striking in both the Doppler shift (black curve in upper right panel) and main peak intensity (red curve), especially in the middle of the displayed part of the observations. The green crosses mark the zero-crossing times of velocity in the shock front passages. At those times the main peak intensity is near its minimum.

The symbols in the lower right panel of Fig.~\ref{shocks} show the intensity of the \ion{Si}{IV} 1403\,\AA\ red-shifted satellite as a function of time. Here we averaged the satellite peak intensities over five consecutive measurements in order to reduce the noise.  One can see that the shock waves leave a weak imprint on the satellite intensity. However, there is an anti-correlation to the intensity of the main component. The satellite's maximum intensity is found at times of mimimum intensity in the main component (see black symbols in the left panels of Fig.~\ref{shocks}, which show the fluctuations averaged over the 9 shock front events having their zero-velocity marked by crosses in the upper right panel, taking the time of vanishing velocity as temporal reference). The error bars in the left panels show the error of the mean value of the displayed signal after averaging over the 9 shock events.

\begin{figure}
    \resizebox{\hsize}{!}{\includegraphics{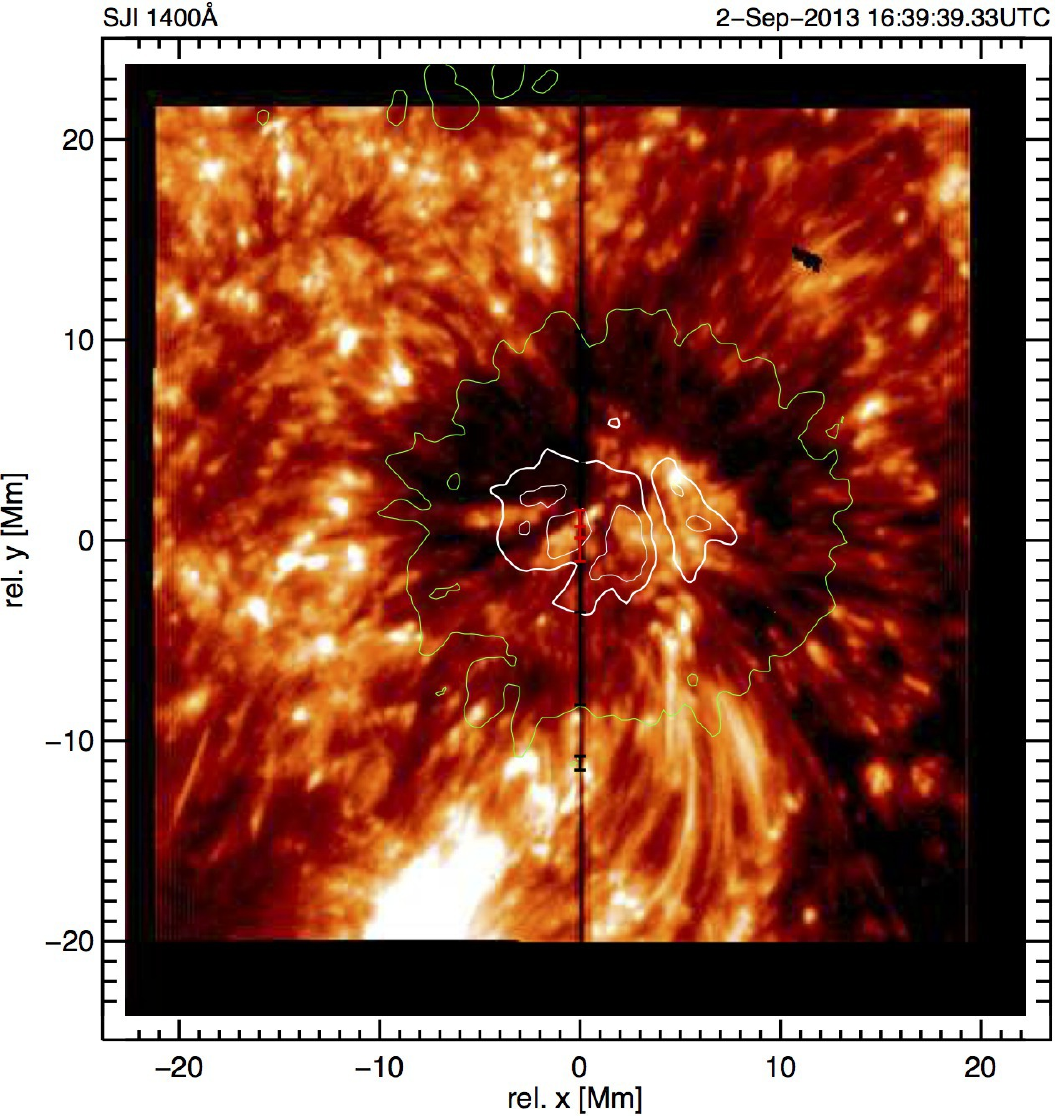}}
     \caption{{IRIS 1400\,\AA\ slit jaw image of the same area of AR\,11836 as the HMI image in the lower left panel of Fig.~\ref{spot}. A time series of these images is available as a movie in the online edition.}}
     \label{movie}
\end{figure}

Interestingly, the satellite intensity (black symbols in the lower left panel of Fig.~\ref{shocks}) is not only anti-correlated to the intensity of the main component, but also shows a different shape. The shock-like shape is much less pronounced in the satellite's intensity profile and reveals a slight preference of opposite sign having a slow rise before and a faster decrease after the peak intensity, which might hint at a downward shock. We will come back to this particular feature further down in the Discussion.

\section{Discussion}

The transition region is highly structured and dynamic \citep[e.g.][]{1992CAS....23.....M}. This fact has become even more obvious in recent high-resolution IRIS observations \citep[e.g.][]{2014Sci...346A.315T,2014Sci...346E.315H}. It is therefore quite remarkable, that the downflow  we observe in a sunspot umbra is rather stationary for nearly 1.5 hours. Has such a phenomenon been observed before? We do believe that the one snapshot obtained during the HRTS I flight by \citet{1982SoPh...77...77D} and \citet{1982SoPh...81..253N} and the SUMER observations of a dual flows in \ion{O}{V} and \ion{N}{V} in sunspots by \citet{2001ApJ...552L..77B} and \citet{2004ApJ...612.1193B} show the same phenomenon, albeit at a different (higher) temperature in the latter cases (\ion{O}{V} and \ion{N}{V} vs \ion{Si}{IV} and \ion{O}{IV}). 

What could be the origin of the feature we see? An obvious candidate, of course, is coronal rain, a commonly observed  phenomenon, in the EUV and even H$\alpha$ \citep[e.g.][]{1996SoPh..166...89W,2001SoPh..198..325S,2005A&A...443..319D,2005ESASP.600E..30M}. In order to obtain information about flows at other temperatures, we inspected co-spatial and co-temporal SDO/AIA data and found material flowing towards the sunspot in all channels. In  the right panel of Fig.~\ref{flows} we show the intensity in the 304\,\AA\  passband, integrated across the path marked by the green line in  the  AIA 304\,\AA\ context image, as a function of the distance from point ``C''. We find plane-of-sky inflows towards the sunspot reaching  velocities of the order  of the (vertical) downflow observed by IRIS (indicated by the continuous line of 90\,km\,s$^{-1}$). One could speculate that these flows feed the downflow plume. However, the signature of the  inflowing material in the  AIA channels disappears before reaching the light-bridge which is marked by the blue dashed lines in the time-distance plot and by the corresponding tick marks in the AIA 304\,\AA\ context image in Fig.~\ref{flows},  leaving the central part of the umbra dominated by the umbral  oscillations. These umbral oscillation are  particularly strong in the  304\,\AA, 171\,\AA, 211\,\AA\ passbands and seem to originate from the center  of the umbra where the satellite  feature is located, as already mentioned above.  

In addition, we inspected the IRIS 1400\AA\ slit-jaw data 
(see movie attached to Fig.~\ref{movie}),  which also shows flows towards the sunspot. Again, as seen in the AIA 304 {\AA} data, the flows become invisible at the outer boundary of the penumbra. No flows or loops can be detected that connect to the core of the umbra.

Since none of these flows apparently connects to the center of the sunspot umbra where we observe the downflow in the transition region, we cannot  ascertain a link between the  supersonic transition region downflow  measured with the IRIS spectrograph and the inflows towards the spot observed with AIA and the IRIS slit-jaw cameras. Moreover, as in most observations of coronal rain, we could only track individual "blobs" falling along the  magnetic field lines.  These blobs, if they really fell through to the satellite region around point ``C'', would appear as strongly intermittent events. The intermittency of coronal rain is also a key characteristic of models of  coronal rain \citep{2003A&A...411..605M,2004A&A...424..289M,2005A&A...436.1067M}. Further, coronal rain is often visible in chromospheric lines such as H$\alpha$ and Ca II, whereas  the downflow we observe is visible only in transition region lines  formed around 10$^5$\,K, with no trace of velocities in the cooler \ion{C}{ii} and \ion{Mg}{ii} lines. We therefore believe that coronal rain is not consistent with most of the observational facts listed in Sec.~\ref{sec:observations}, but we cannot exclude it with certainty.

Another possible explanation, which we find more appealing, in particular considering the remarkable stability of the supersonic downflow, is a siphon flow, along the lines discussed by e.g. \cite{1980SoPh...65..251C} and \citet{1981SoPh...69...63N}, who studied stationary flow solutions in a coronal magnetic arch. In the following, we will refer to Noci's paper since it makes predictions that are more relevant to our observations. His ``class (vii)'' solutions are of particular interest here. These solutions describe subsonic-supersonic steady flows with stationary shocks which adjust the flow to the boundary conditions in the second foot-point. The location and strength of the stationary shock is determined by the pressure ratio between the two foot-points. With decreasing pressure in the downflow foot-point the shock moves closer towards the foot-point with increasing jumps in temperature and density. We should point out, though, that Noci's solutions are for coronal loops at coronal temperatures ($>$ 1\,MK), whereas we observe the flow at transition region temperatures. However, we see no reason why these solutions should be limited to coronal temperatures and speculate that the supersonic downflow we see at transition region temperatures is the signature of a stationary shock in a siphon flow close to the foot-point in the sunspot umbra.

In this picture, the main component dominated by the umbral 3-min shock waves is formed in the part of the transition region underneath the stationary shock of the siphon flow. As mentioned above, the strength of the stationary shock in Noci's solutions depends on the pressure ratio between the foot-points. Due to the strong non-linear umbral 3-min oscillations, one cannot expect a constant pressure ratio between the two foot-points. One might therefore expect a modulation of the stationary shock intensity by the underlying umbral 3-min shock waves. At minimum intensity in the main component, which coincides with minimum pressure and minimum temperature at the base of the downflow foot-point, Noci's solutions would predict the strongest stationary shock in the siphon flow. This is exactly what we observe. As can be seen in the lower left panel of Fig.~\ref{shocks},  the maximum intensity of the supersonic downflow coincides with minimum intensity of the main component.

\begin{figure}
    \resizebox{\hsize}{!}{\includegraphics{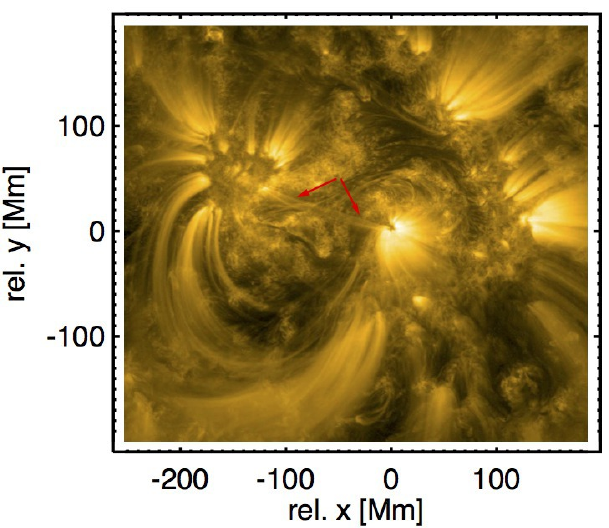}}
     \caption{AIA 171\,\AA\ image of the same area as in Fig.~\ref{spot} taken a day later on Sep 3, 2013 14:20 UT. Note the coronal loops (indicated by red arrows) connecting the dark quadrant of the spot directly to the trailing plage of opposite polarity.}
     \label{siphon}
\end{figure}

Without having identified the upflow foot-point, we are not in a position to conclusively claim the detection of a siphon flow at transition region temperatures; we can merely speculate. Along these lines it is interesting to note that about 21 hours after the observations presented here, the AIA 171\,\AA\ channel shows prominent coronal loops connecting the dark quarter of the sunspot umbra directly to the trailing plage of opposite polarity (Fig.~\ref{siphon}). Could there have been a similar loop at cooler, transition region temperatures during our observations?

In order to get some further insight into the evolution and extent of the downflow, we inspected all IRIS observations of AR\,11836 on 2015 September 2 and 3. In addition to the data set studied in detail in this paper, there are 7 more data sets of AR\,11836:  5 large coarse 64-step rasters with 2\arcsec\ step width, and 2 two-step sparse rasters. One of these data sets, the large coarse raster starting on 2013-09-02T11:56:35 indeed shows clear signatures of a supersonic downflow for 4 steps ($\#$35-38) during all 5 raster scans. However, this downflow occurs at a different location in the sunspot, extending  from the southern part of the most prominent light-bridge visible in Fig.\,\ref{spot} (x $\approx$ 4\,Mm, y $\approx$ -2\,Mm) into the penumbra to the West. The measured downflow speed is smaller ($\approx$ 75 km\,s$^{-1}$). Since there are only 5 scans (with a scan cycle time of 188\,s), it is difficult to infer information about temporal variations and its relation to the shock-wave dominated main component. All we can say is that the feature lasted for more than 12\,min, has a similar extent along the slit as the downflow studied here and extended over at least 8\arcsec\ in the direction perpendicular to the slit. The large coarse raster scan following the observations studied in this paper (starting at 18:29:35 UT) also shows some hints of a downflow, but with a much weaker signal and not in consecutive frames, as if it had split up. None of the other four data sets from September 2--3 shows any signs of a supersonic downflow. In a future paper, in which we will study the range of main properties, locations and rate of occurrence of this phenomenon, we will analyse these data sets in more detail.

From the measured density (cf.~Section \ref{sec:plasma:density}) of the downflow and its speed, $v \approx 90$\,km\,s$^{-1}$, we can estimate the mass flux as $\rho\, v = N_\mathrm{e}\, m_\mathrm{p}\, (N_\mathrm{H}/N_\mathrm{e})\, v$, where $\rho$ is the mass density and $m_\mathrm{p}$ is the proton mass, which yields $\approx 5 \times 10^{-7}\,\mathrm{g}\,\mathrm{cm}^{-2}\,\mathrm{s}^{-1}$\, assuming $N_\mathrm{H}/N_\mathrm{e}=0.83$. This is a substantial flux. It would evacuate the overlying corona on time scales of the order of 10\,s. Even a prominence (column number density of $1 \times 10^{19}\,\mathrm{cm}^{-2}$) would be drained in about 40\,s. For the interpretation of the flow as a siphon flow, such a high mass flux does not pose a problem. 

\section{Conclusions}

We have observed a small-scale, supersonic downflow of about 90\,km\,s$^{-1}$  in the transition region above the sunspot umbra of AR\,11836. The downflow is not visible in the chromospheric lines, which only show an intensity enhancement at the location of the downflow. The downflow shows up as red-shifted, well-separated ``satellite'' lines of the \ion{Si}{IV} and \ion{O}{IV} transition region lines and is remarkably steady over the observing period of nearly 80\,min. The satellite lines do not participate in the 3-min shock wave Doppler maneuvres of the main component, but show a weak anti-correlated signature of the 3-min oscillations in intensity. 
We estimate the mass flux in the stationary flow to be $\approx 5 \times 10^{-7}\,\mathrm{g}\,\mathrm{cm}^{-2}\,\mathrm{s}^{-1}$. This mass flux is relevant for the atmospheric structure, as it would evacuate an overlying quiet coronal column in about 10\,s.

We interpret these findings as evidence of  a stationary termination shock of a supersonic siphon flow in a cool loop rooted in the central umbra of the spot, and we speculate that the supersonic flows seen by \citet{1982SoPh...77...77D}, \citet{1982SoPh...81..253N},  \citet{2001ApJ...552L..77B}, and \citet{2004ApJ...612.1193B} are caused by the same phenomenon. We surmise that the 3-min shock waves visible in the main component of the lines form in the transition region underneath the stationary shock and represent a perturbation to the boundary condition of the stationary flow. This would explain the anti-correlation between the intensity of the main and the red-shifted satellite components. 

Models of these supersonic siphon flows with termination shocks were developed by \citet{1980SoPh...65..251C} and \citet{1981SoPh...69...63N} for hot, coronal loops, although they should be applicable to transition region temperatures as well.  There are numerical studies on siphon flows  in coronal loops \citep[e.g.:][]{1988ApJ...334..489M,1991ApJ...382..338S,1995A&A...294..861O}, some including shocks \cite[e.g][]{1995A&A...300..549O}, but we are not aware of any numerical studies reproducing shocks of the magnitude we are observing at transition region temperatures. Numerical studies of flows in the transition region \cite[e.g.][]{1984ApJ...280..416A,2004ApJ...617L..85P,2006ApJ...638.1086P,2010ApJ...718.1070H} were usually targeted at reproducing the average red-shifts observed at those temperatures, and thus do not predict strong flows and shocks. It would therefore be interesting to consider numerical studies of solutions of the class studied by Noci, but for cool loops that never reach coronal temperatures.

There are numerous observations in the photosphere, chromosphere and corona that have been interpreted as siphon flows  \citep[e.g.][]{1992A&A...261L..21R,2000A&A...359..716S,2006ApJ...645..776U,2010AN....331..574B,2012A&A...537A.130B}. What are their relations, if any, to the supersonic downflows measured in the transition region over sunspot umbrae? 
Observations should also clarify whether steady-state supersonic downflows in the transition region are a common phenomenon, visible only in certain circumstances, or whether they are a rather rare phenomenon. If they are a regularly occurring phenomenon, what is the range of their main properties? We find it intriguing, that the events with a single, clearly separated red-shifted component such as those observed by \citet{1982SoPh...77...77D}, \citet{1982SoPh...81..253N}, one of the events of \citet{2001ApJ...552L..77B}, and this work, have approximately the same speed ($\approx$\,90\,km\,s$^{-1}$), which is at the high end of the average downflow speeds measured for 5 sunspots by \citet{2004ApJ...612.1193B}. 

\begin{acknowledgements}
IRIS is a NASA small explorer mission developed and operated by LMSAL with mission operations executed at NASA Ames Research center and major contributions to downlink communications funded by ESA and the Norwegian Space Center (NSC). 
CHIANTI is a collaborative project involving George Mason University, the University of Michigan (USA) and the University of Cambridge (UK).
We gratefully acknowledge helpful discussions with Bart De Pontieu, Joe Gurman, Jim Klimchuk, Daniele Spadaro, Han Uitenbroek. This work has also benefited from discussions at 
the International Space Science Institute (ISSI) meeting on 
``Heating of the magnetized chromosphere'' from 5-8 January, 2015, 
where many aspects of this paper were discussed with other colleagues. We also thank an anonymous referee for useful comments.
\end{acknowledgements}

\bibliographystyle{aa} 

\end{document}